\documentclass[
    aps,12pt,final,notitlepage,
    oneside,onecolumn,nofootinbib,noshowpacs,
    nolongbibliography,centertags
]{revtex4-2}

\usepackage[english]{babel}

\usepackage[]{graphicx}
\graphicspath{{figures/}}

\usepackage{natbib}

\usepackage[usenames,dvipsnames]{xcolor} % link colors tweak
\usepackage[
    colorlinks = true,
    urlcolor = MidnightBlue,
    linkcolor = Red,
    citecolor = NavyBlue
]{hyperref}

\newcommand{{\yeas}}{{YEASa}}

\newcommand{{\ethr}}{\varepsilon_{\text{thr.}}} % threshold energy
\newcommand{{\lnA}}{\langle\ln{A}\rangle} % <lnA>
\newcommand{{\xmax}}{x_{\text{max}}}      % Xmax
\newcommand{{\Xmp}}{x_{\text{max}}^p}
\newcommand{{\XmFe}}{x_{\text{max}}^{\text{Fe}}}
\newcommand{{\XmExp}}{x_{\text{max}}^{\text{exp}}}
\newcommand{{\rhom}}{\rho_{\mu}}          % muon density
\newcommand{{\rhos}}{\rho_s}              % surfase density

\newcommand{\Nexp}{N_{\mu}^{\text{exp}}}
\newcommand{\Np}{N_{\mu}^p}
\newcommand{\NFe}{N_{\mu}^{\text{Fe}}}

\newcommand{{\qgs}}{{QGSJet}}
\newcommand{{\epos}}{{EPOS}}
\newcommand{{\eposlhc}}{{EPOS-LHC}}
\newcommand{{\sibyll}}{{Sibyll}}
\newcommand{{\fluka}}{{FLUKA}}
\newcommand{\corsika}{{CORSIKA}}

\newcommand{{\usec}}{{~$\mu$s}}    % usec
\newcommand{{\sqrm}}{{~m$^2$}}     % m^2
\newcommand{{\sqrkm}}{{~km$^2$}}   % km^2
\newcommand{\depth}{{~g/cm$^2$}}   % g/cm^2
\newcommand{\degr}{^{\circ}}     % degrees

\begin{document}

\title{Status of the Yakutsk air shower array and future plans}

\author{A.\,K.~Alekseev}

\author{E.\,A.~Atlasov}

\author{N.\,G.~Bolotnikov}

\author{A.\,V.~Bosikov}

\author{N.\,A.~Dyachkovskiy}

\author{N.\,S.~Gerasimova}

\author{A.\,V.~Glushkov}

\author{A.\,A.~Ivanov}

\author{O.\,N.~Ivanov}

\author{D.\,N.~Kardashevsky}

\author{I.\,A.~Kellarev}

\author{S.\,P.~Knurenko}

\author{A.\,D.~Krasilnikov}

\author{A.\,N.~Krivenkov}

\author{I.\,V.~Ksenofontov}

\author{L.\,T.~Ksenofontov}

\author{K.\,G.~Lebedev}

\author{S.\,V.~Matarkin}

\author{V.\,P.~Mokhnachevskaya}

\author{E.\,V.~Nikolaeva}

\author{N.\,I.~Neustroev}

\author{I.\,S.~Petrov}

\author{N.\,D.~Platonov}

\author{A.\,S.~Proshutinsky}

\author{A.\,V.~Saburov}
\email{tema@ikfia.ysn.ru}

\author{I.\,Ye.~Sleptsov}

\author{G.\,G.~Struchkov}

\author{L.\,V.~Timofeev}

\author{B.\,B.~Yakovlev}

\affiliation{
    Yu.~G.~Shafer Institute of Cosmophysical Research and Aeronomy SB RAS, \\
    31 Lenin Ave., 677027 Yakutsk, Russia
}

%\date{}

\begin{abstract}
    The Yakutsk Extensive Air Shower Array has been continuously operating for
    more than 50 years (since 1970) and up until recently it has been one of
    world's largest ground-based instruments aimed at studying the properties
    of cosmic rays in the ultra-high energy domain.  In this report we discuss
    results recently obtained at the array~--- on cosmic rays energy spectrum,
    mass composition and directional anisotropy~--- and how they fit into the
    world data. Special attention is paid to the measurements of muonic
    component of extensive air showers.  Theoretical results of particle
    acceleration at shocks are also briefly reviewed. Future scientific and
    engineering plans on the array modernization are discussed.
\end{abstract}

\maketitle

\section{Introduction}
\label{sect:intro}

The only way to study the properties of cosmic rays (CR) with energies above
$\sim 10^{15}$~eV is by registering extensive air showers (EAS)~---
cascades of secondary particles produced by high-energy CRs hitting Earth's
atmosphere. On one hand by studying the EAS properties one can investigate
hadron interactions at energies unreachable for terrestrial particle
acceleration technology. On the other hand~--- EAS method provides the basis
for obtaining crucial information on primary CR properties: their energy
spectrum, mass composition and distribution of arrival directions on the
celestial sphere. Hence, EAS study is a dedicated area of CR physics at the
intersection of astrophysics and nuclear physics.

There are many techniques of EAS registration and all of them are based on the
detection of shower particles and accompanying emission in various ranges. The
most common type of an instrument intended for EAS registration is a
ground-based EAS array which is, basically, a network of particle detectors
distributed over a certain area of ground surface.

After the prediction was made about CR spectrum abrupt cut-off at
$\simeq 6 \times 10^{19}$~eV due to CR interaction with recently
discovered cosmic microwave background (CMB)~\cite{Zatsepin:jetpl(1966),
Greizen:prl(1966)}, a series of projects had been started around the world,
aiming at creation of large EAS arrays capable of providing sufficient
statistics in the ultra-high energy domain. One of this projects have resulted
in the creation of the Yakutsk extensive air shower array (\yeas).

The {\yeas} is a stationary research site located in the Lena river valley
55~km south of Yakutsk (Russia) (61.7$\degr$\,N, 129.4$\degr$\,E,
$\simeq 100$~m above the sea level, atmospheric depth~--- 1020\depth).
It is one of the longest running experiments that register the flux of the
ultra-high energy CRs (UHECR): it has been continuously operating since 1973.
From the very beginning it was created as a complex instrument, capable of
measuring several EAS components with different types of detectors. Charged and
electromagnetic (e-m) components of air showers are measured with surface
scintillation detectors (SD). Muons are registered with the similar detectors
buried below the ground level: such placement prevents the contamination of the
muon data with e-m component by providing a shielding with energy threshold
$\simeq 1$~GeV.

The array controls a wide range of CR energy ($E$), from $10^{15}$~eV up to and
above $10^{19}$~eV, thanks to its three-threshold trigger. The basis of the
array is a network of forty-nine observation detector points (or ``stations'')
that form a triangular pattern and covers the area of 11{\sqrkm}. Each triangle
serves as a local trigger for events selection. The so-called ``Trigger-500''
(T500) is formed by triangles with 500-m side and selects events with $E \simeq
10^{17}$~eV. The ``Trigger-1000'' (T1000) is formed by triangles with 1000-m
side and selects events with energy starting from about $10^{18}$~eV. Each
station contains a pair of 2{\sqrm} scintillation detectors operating in
2{\usec} coincidence mode. Scintillation detectors are continuously calibrated
by array's data acquisition and synchronization system (DAQ) and are adjusted
by amplitude density spectra of background cosmic
muons~\cite{Glushkov:yafsoan(1974), Borodina:yafsoan(1974)}.

Nineteen stations in central part of the array are equipped with integral
optical detectors providing measurement of Cherenkov light emission produced by
charged component of air showers (Main ChA). In the center of the array are
located detectors that form a standalone small array that selects events by
Cherenkov light with threshold energy around $10^{15}$~eV~--- the Small
Cherenkov Array (Small ChA)~\cite{Knurenko:sciedu(1998)}. It has a rather
complex trigger that accounts for various combinations of fired detectors. The
latest addition to this setup is a system of three observational points, each
housing fast differential Cherenkov detectors based on camera-obscura principle
(the ``Obscura'')~\cite{Egorov:pos(2017)}. First point, ``Obscura-1'', also
contains a set of scintillation detectors and muon
telescopes~\cite{Knurenko:prd(2020)}. The ``Obscura'' system is also equipped
with a DAQ that controls a set of dipole antennas registering radio-emission
of EAS within the 30-35~MHz frequency band~\cite{Knurenko:nima(2017)}.

The array has been continuously operating for more than 50~years. During
this period an extensive and uniform data set has been accumulated. Here we
give an overview of recent results obtained with this instrument.

\section{Cosmic Ray Energy Spectrum}

The main EAS parameters at the {\yeas}~--- arrival direction, location of a
showers core and primary energy $E$~--- are reconstructed by the lateral
distribution function (LDF) of e-m and charged components registered with SD.
This particles travel through a multi-layered shield with a total thickness
$\simeq 2.5${\depth} and through a 5~cm scintillator, releasing a
portion of energy $\Delta E_s$ measured in arbitrary units $\rhos(r)$, which is
proportional to the $E_1 = 11.75$~MeV, the vertical muon equivalent (VEM).

The LDFs of the SD response were obtained within the frameworks of four hadron
interaction models for primary protons and iron nuclei for energy range
$10^{17} - 10^{19.5}$~eV~\cite{Glushkov:yadf(2018)}. The main energy estimator
at the {\yeas} is $\rhos(600)$~--- a measured particle density at 600~m from
the shower core. The energy is reconstructed with the use of quasi-calorimetric
method based on estimation of the energy fraction dissipated by e-m component
in the atmosphere described in~\cite{Egorov:tokyo(1993)}. EAS simulations
performed with {\corsika} code~\cite{corsika} gave a new refined estimation of
CR energy:

\begin{figure}[!htbp]
    \centering
    \includegraphics[width=0.95\textwidth]{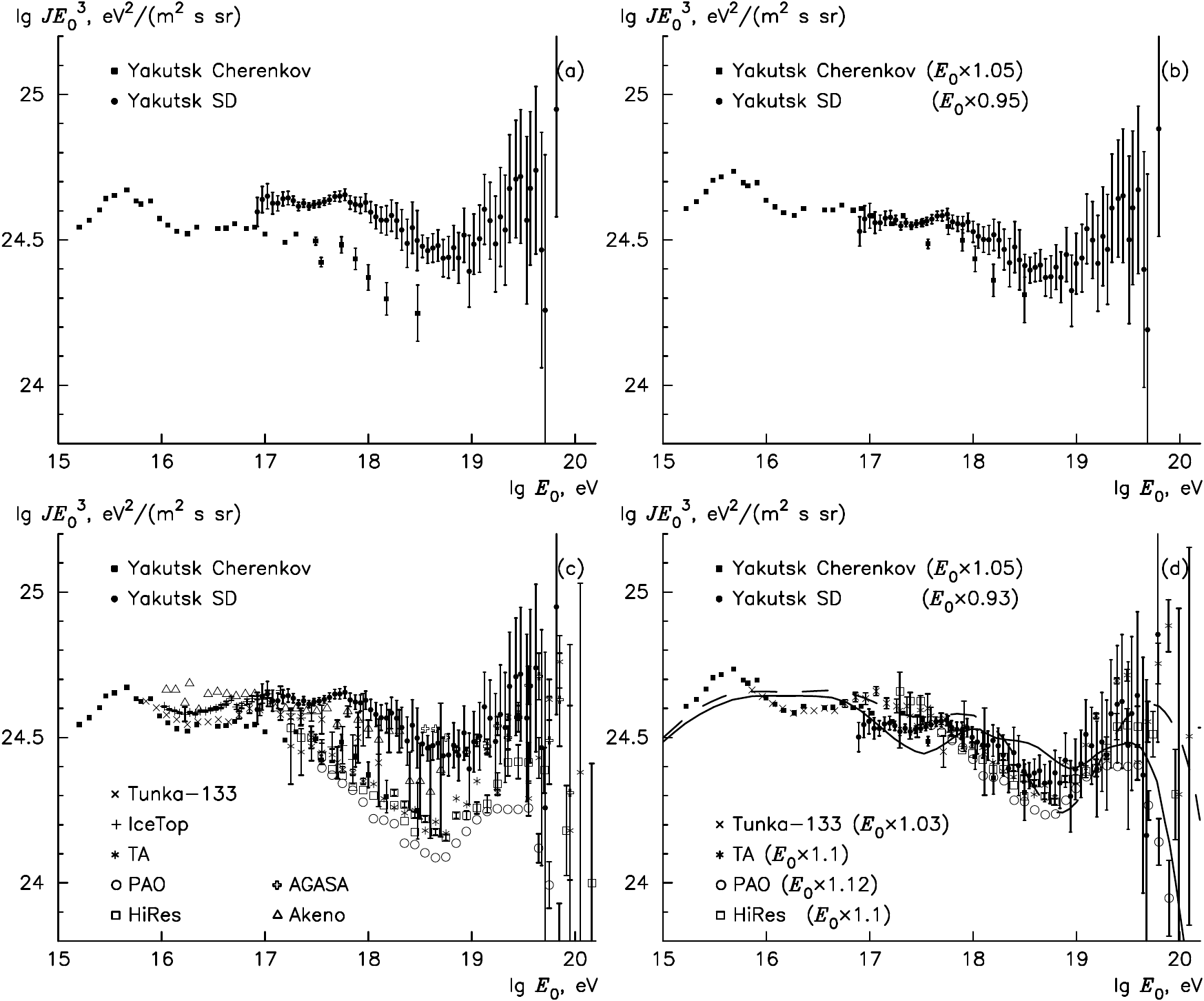}
    \caption{
        (a)\,Energy spectrum of CR measured by the {\yeas}: black circles
        represent the data from the main SD grid, black squares~---
        measurements with Cherenkov detectors.
        (b)\,Same as in panel (a) with applied $\pm 5$\,\% corrections to both
        estimates.
        (c)\,A comparison between the CR spectra measured by different world
        experiments: Tunka-133~\cite{Berezhnev:nima(2012)},
        IceCube~\cite{Aartsen:prd(2013)}, TA~\cite{Abbasi:app(2016)},
        PAO~\cite{Aab:icrc2015}, HiRes~\cite{Zundel:pos(2016)},
        Akeno~\cite{Nagano:gnucl(1992)} and AGASA~\cite{Shinozaki:npb(2006)}.
        (d)\,Same as in panel (c) but with energy scale of each experiment
        tuned by its individual coefficient. Lines represent the predictions of
        the CR diffusive shock acceleration model~\cite{Berezhko:app(2012)}.
    }
    \label{fig:1}
\end{figure}

\begin{equation}
    E = (3.76 \pm 0.3) \times 10^{17} (\rhos(600,0\degr))^{1.02 \pm
    0.02}\text{,}
    \label{eq:EPrim}
\end{equation}
where $\rhos(600,0{\degr})$ is a $\rhos(600)$ value converted to vertical
shower direction. The resulting energy spectrum is shown on Fig.~\ref{fig:1}(a)
with black circles (Yakutsk SD). It includes more than $10^{6}$ events
registered at the {\yeas} during the continuous period of operation from 1974
to 2017.  Simulations gave energy estimation lower by factor $\simeq
1.28$ compared to the previous one, given in~\cite{Egorov:tokyo(1993)}.

With black squares on the same panel is shown energy spectrum obtained by the
measurement of Cherenkov light emission at the {\yeas}~--- with both Main ChA
and Small ChA (Yakutsk Cherenkov)~\cite{Knurenko:epj(2013)}. Energy was
estimated with the use of a similar quasi-calorimetric technique that utilizes
a flux of Cherenkov photons at a certain core distance $Q(r)$ as an energy
estimator:

\begin{equation}
    E = (9.12 \pm 2.28) \times 10^{16} \left(\frac{Q(150)}{10^7}\right)^{0.99
    \pm 0.02}\text{,}
    \label{eq:EPrimQs}
\end{equation}

\begin{equation}
    E = (8.91 \pm 1.96) \times 10^{17} \left(\frac{Q(400)}{10^7}\right)^{1.03
    \pm 0.02}\text{,}
    \label{eq:EPrimQb}
\end{equation}
where $Q(150)$ and $Q(400)$ are energy estimators at 150~m and 400~m from
shower core correspondingly. Expression (\ref{eq:EPrimQs}) is used in energy
range $(5-500)$~PeV and expression (\ref{eq:EPrimQb})~--- to estimate primary
energy of showers above $5 \times 10^{17}$~eV.

At first glance it follows from Fig.~\ref{fig:1}(a) that the spectrum measured
with the use of Cherenkov technique is significantly lower than the measured by
SD. But given that the uncertainties of calorimetric methods (\ref{eq:EPrim})
and (\ref{eq:EPrimQs}-\ref{eq:EPrimQb}) are above 20\,\% and about 15\,\%
correspondingly, if one applies just a 5\,\% correction to the estimated energy
in both spectra, then they start to agree within experimental errors, as shown
on Fig.~\ref{fig:1}(b).

On lower panels of Fig.~\ref{fig:1} the results of the {\yeas} are shown in
comparison with CR spectra measured by different world experiments:
Tunka-133~\cite{Berezhnev:nima(2012)}, IceCube~\cite{Aartsen:prd(2013)},
Telescope Array (TA)~\cite{Abbasi:app(2016)}, the Pierre Auger Observatory
(PAO)~\cite{Aab:icrc2015}, HiRes~\cite{Zundel:pos(2016)},
Akeno~\cite{Nagano:gnucl(1992)} and AGASA~\cite{Shinozaki:npb(2006)}. On
Fig.~\ref{fig:1}(c) values are shown without any correction of energy scale, on
Fig.~\ref{fig:1}(d) energy scale of each experiment was tuned by its own
coefficient, also shown on panel (d). Lines represent the predictions of the
CR diffusive shock acceleration model~\cite{Berezhko:app(2012)} described in
details in section~\ref{sec:theory}. It is seen that after applying such energy
rescaling all results agree with each other within experimental errors.

\section{Cosmic Ray Mass Composition}

\begin{figure}[!htbp]
    \centering
    \includegraphics[width=0.95\textwidth]{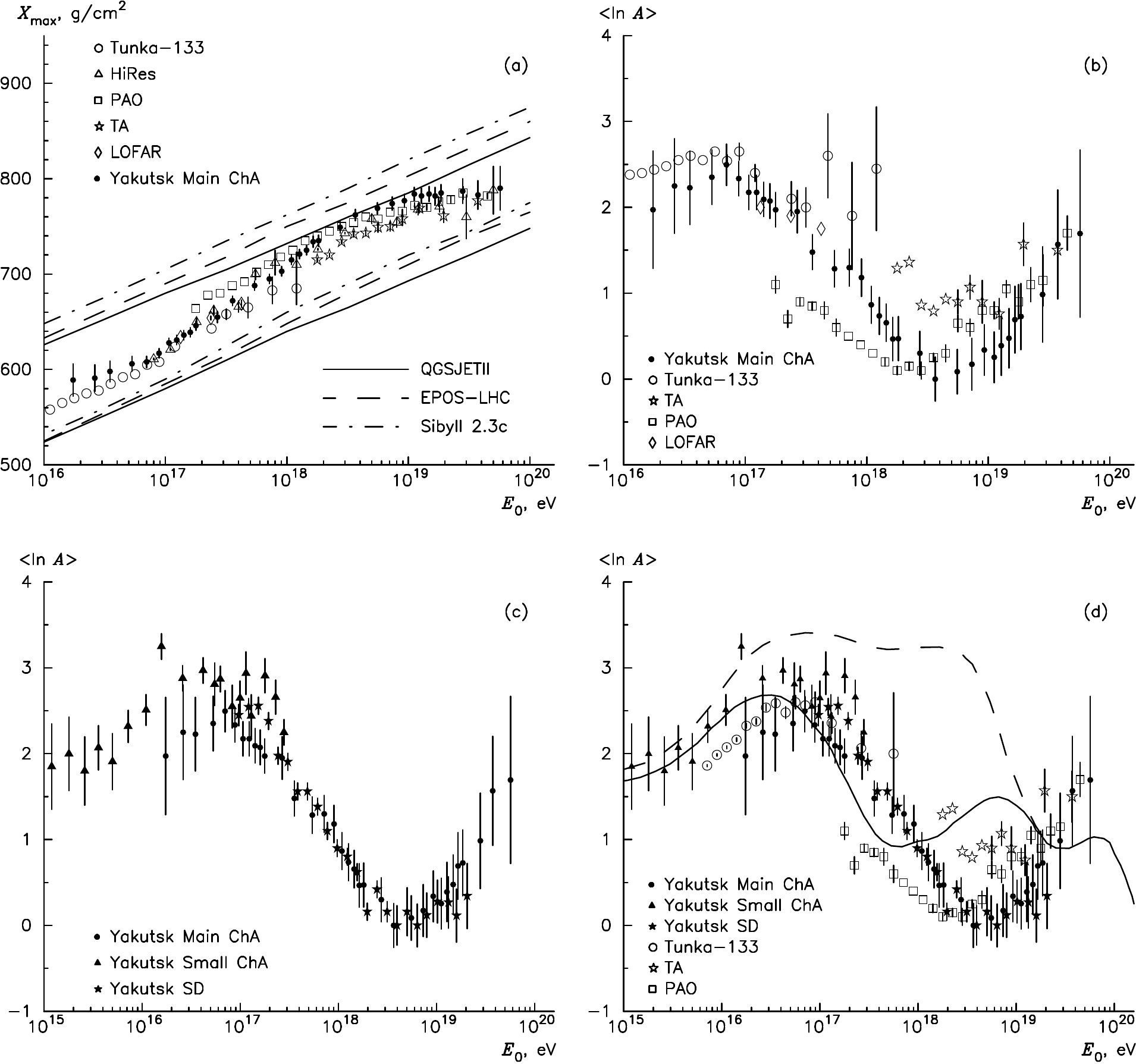}
    \caption{
        (a)\,The depth of maximum air shower development $\xmax$ reconstructed
        from LDF of Cherenkov light measured at the
        {\yeas}~\cite{Knurenko:asr(2019)} (black circles). Also are shown the
        results of Tunka-133~\cite{Prosin:epj(2016)} (empty circles),
        HiRes~\cite{Ulrich:app(2012)} (empty triangles),
        PAO~\cite{Bellido:pos(2018)} (empty squares),
        TA~\cite{Abbasi:apj(2018)} (stars) and
        LOFAR~\cite{Buitnik:nature(2016)} (empty diamonds). Lines represent the
        $\xmax$ energy dependencies according to the predictions of three
        hadron interaction models.
        (b)\,Mean logarithm of the CR atomic weight $\lnA$ derived from the
        $\xmax$ data shown in panel (a), symbols are the same.
        (c)\,A comparison of the $\lnA$ estimations obtained with the {\yeas}'s
        SD (black stars)~\cite{Pravdin:pazh(2018)} and Cherenkov detectors:
        black circles~--- measurements with the Main
        ChA~\cite{Knurenko:asr(2019)}, black triangles~--- with the Small
        ChA~\cite{Knurenko:epj(2013)}.
        (d)\,A comparison of Yakutsk results with the world data. Lines
        represent the predictions of the CR diffusive shock acceleration model
        for two scenarios: the ``dip''-scenario (solid line) and the
        ``ankle''-scenario (dashed line)~\cite{Berezhko:app(2012)}.
    }
    \label{fig:2}
\end{figure}

The {\yeas} has been continuously registering CRs at energy above
$10^{17}$~eV~--- for 50 years. During this period a unique dataset have been
accumulated on e-m, muon and Cherenkov components of EAS. These measurements
allowed us to reconstruct the cascade curve of shower development and its
parameters, namely~--- the depth of maximum shower development ($\xmax$) which
is sensitive to CR mass composition and further investigate the dynamics of
its change with increase of energy. The $\xmax$ was reconstructed from
LDF of Cherenkov light in EAS measured at the
{\yeas}~\cite{Knurenko:asr(2019)}. The results are shown on
Fig.~\ref{fig:2}(a) with black circles (Main ChA) together with measurements of
different arrays: Tunka-133~\cite{Prosin:epj(2016)},
HiRes~\cite{Ulrich:app(2012)}, PAO~\cite{Bellido:pos(2018)},
TA~\cite{Abbasi:apj(2018)} and LOFAR~\cite{Buitnik:nature(2016)}. With lines
are denoted $\xmax$ energy dependencies according to the predictions of hadron
interaction models {\qgs}~II-04~\cite{Ostapchenko:prd(2011)},
\eposlhc~\cite{Pierog:prc(2015)}, and \sibyll-2.3c~\cite{Riehn:pos(2018)}.

The mean atomic number of primary CR particles ($\lnA$) can be estimated from a
relation that follows from principle of nucleon superposition as:

\begin{equation}
    \lnA = \frac{\Xmp - \XmExp}{\Xmp - \XmFe} \ln{56}\text{,}
    \label{eq:lnA}
\end{equation}
where values of $\xmax$ were obtained in experiment (``exp'') and in model
calculations for primary protons ($p$) and iron nuclei (Fe). Interpretation of
$\xmax$ measurements is shown on Fig.~\ref{fig:2}(b). This result was obtained
for limited data selection with \qgs~II-04 model, which explains discrepancy
with the shown data.

Another method of the CR mass composition estimation utilized at the
{\yeas}~--- is to derive $\lnA$ values from parameters of LDF measured by
SD~\cite{Pravdin:pazh(2018)}:

\begin{equation}
    \lnA = w_p \ln{1} + w_{\text{Fe}}\ln{56}\text{,}
    \label{eq:lnASD}
\end{equation}
where $w_p = 1 - w_{\text{Fe}}$ and $w_{\text{Fe}} = \lnA / \ln{56}$. This
approach gives us:

\begin{equation}
    w_{\text{Fe}} = \frac{d_{\text{exp}} - d_p}{d_{\text{Fe}} - d_p}\text{,}
\end{equation}
where $d$ is a parameter of the SD LDF local steepness. On Fig.~\ref{fig:2}(c)
is shown a comparison of $\lnA$ estimations obtained with two methods described
above. Black circles denote results obtained with Main
ChA~\cite{Knurenko:asr(2019)}; with black triangles~--- results obtained with
Small ChA~\cite{Knurenko:epj(2013)}; stars~--- estimation with the use of SD
technique described in~\cite{Pravdin:pazh(2018)}.

Comparison of Yakutsk results with the world data is shown on
Fig.~\ref{fig:2}(d). Lines represent the predictions of the diffusive CR
shock acceleration model within two scenarios described in details in
Section~\ref{sec:theory}. As can be seen from the figure, there are
irregularities of the value of $\lnA$, due to a change in mass composition. It
was found that at energies up to $10^{18}$~eV CR composition is abundant with
nuclei with atomic numbers in the $4-56$ region. In the $10^{18}-10^{19}$~eV
interval the fraction of protons starts increasing and peaks at $50-80$\,\%. At
energy above $10^{19}$~eV CR mainly consist of helium nuclei, nuclei from the
CNO group and heavier elements~\cite{Knurenko:asr(2019), Pravdin:pazh(2018)}.

\begin{figure}[!htbp]
    \centering
    \includegraphics[width=0.75\textwidth]{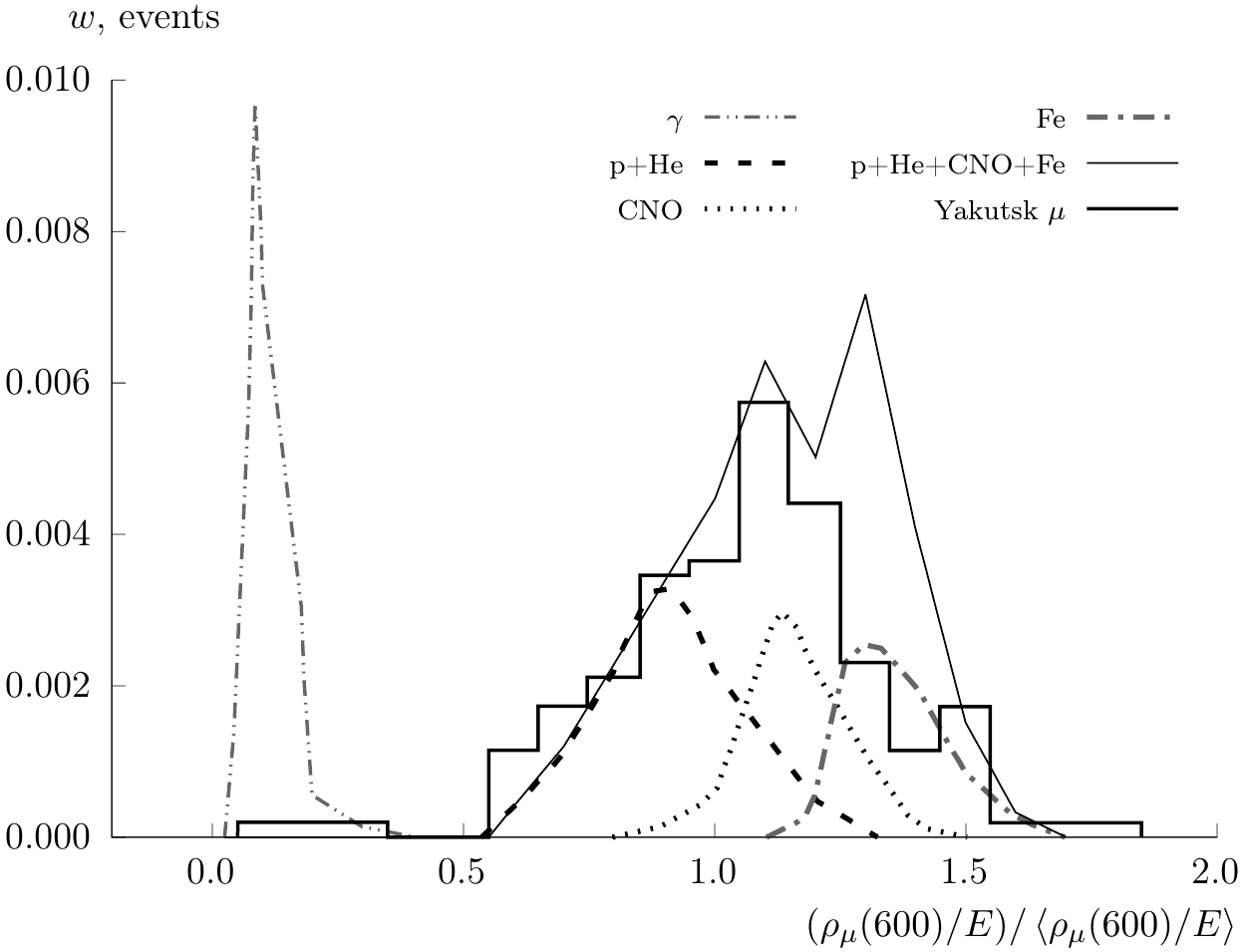}
    \caption{
        Distribution of the muon numbers in showers with energy $(5-50)$~EeV
        and zenith angles $\theta \leq 60 \degr$ by the {\yeas}
        data~\cite{Knurenko:prd(2020)}. Data are normalized to vertical
        direction ($\theta = 0\degr$). Dashed curve denotes $p$ + He, dots
        denote CNO, and dash-dotted curve denotes Fe. The solid curve indicates
        the total value of $p$ + He + CNO + Fe nuclei.
    }
    \label{fig:3}
\end{figure}

On Fig~\ref{fig:3} an estimation of the CR mass composition is shown, obtained
with a slightly different technique: by combining measurements performed on the
main muon setup of the {\yeas} and on muon telescopes included in the Small ChA
setup. We have estimated the CR mass composition in energy range $(5
\times 10^{18} - 5 \times 10^{19})$~eV by muon fraction in registered EASs. It
was found that $50 \pm 10$\,\% of CRs in this energy region are protons and
helium nuclei, $32\pm 6$\,\%~--- nuclei from the CNO group, $16\pm 8$\,\%~---
iron nuclei and about $2$\,\% are primary gamma-photons of
UHE~\cite{Knurenko:prd(2020)}. This preliminary result was obtained within the
framework of \qgs~II-04 model~\cite{Ostapchenko:prd(2011)} and is roughly
consistent with results shown on Fig.~\ref{fig:2}.

\subsection{\it Upper limit of the integral photon flux in the ultra-high
energy domain}

\begin{figure}[!htbp]
    \centering
    \includegraphics[width=0.75\textwidth]{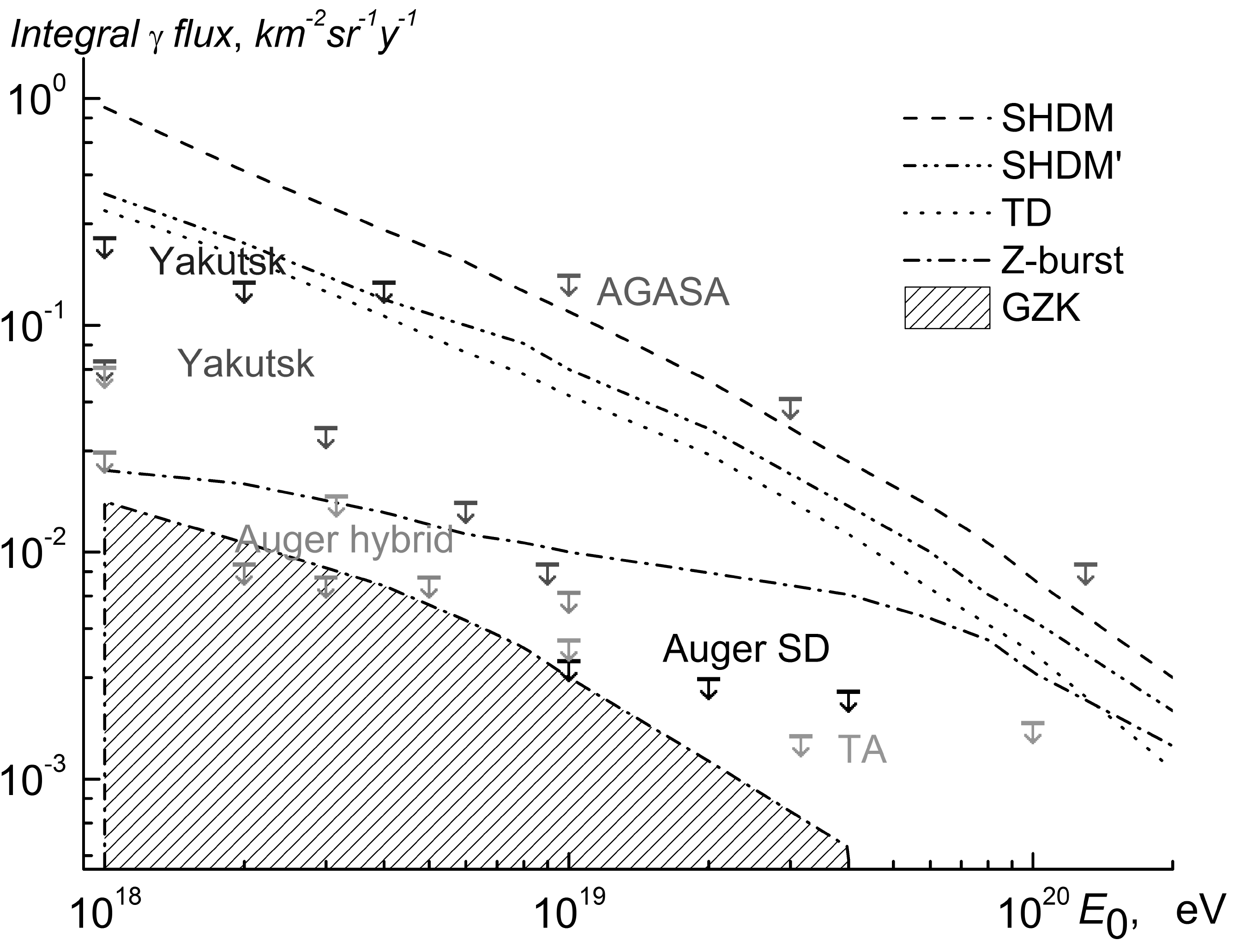}
    \caption{
        Integral flux of primary photons of cosmic rays according to the
        {\yeas} data~\cite{Knurenko:jetpl(2018)}, with previous
        data~\cite{Glushkov:prd(2010)} compared to PAO~\cite{Abraham:app(2008),
        Aab:jcap(2017)}, AGASA~\cite{Shinozaki:apj(2002)}, and Telescope
        Array~\cite{Rubtsov:pos(2018)} experiments and ``top-down'' predictions
        for sources: dashed line represents super heavy dark matter
        (SHDM)~\cite{Ellis:prd(2006)}, dash-dot-dot line represents
        SHDM'~\cite{Bhattacharjee:pr(2000), Berezinsky:plb(1969)}, dotted line
        represents topological defects~\cite{Bhattacharjee:pr(2000),
        Berezinsky:plb(1969)}, dash-dotted line represents
        Z-bursts~\cite{Bhattacharjee:pr(2000), Berezinsky:plb(1969)}, and
        shaded region represents GZK mechanism of gamma-ray photons generation
        with extremely high energies~\cite{Kalashev:jetp(2008)}.
    }
    \label{fig:4}
\end{figure}

Photons do not deviate from their initial direction in magnetic fields and
because of this property are a good instrument for the search of point sources
of UHECRs. That is why any information on primary gamma-photons can be used for
deriving the properties of the CR sources and for studying the interaction of
UHE photons with photonic field of the Universe. The development of EAS
initiated by neutral particles should significantly differ from that in EAS of
hadronic nature. Neutral particles generate air showers with $\xmax$ deep in
the atmosphere; and EASs from primary gamma-photons contain significantly less
muons since they are dominated by e-m component.

A multi-component experimental data set of the {\yeas} allowed us to select
showers whose properties were hinting at their photonic origin. The final
selection included 11 candidate events. The performed analysis resulted in
estimation of the upper limit for integral primary photon flux at energy above
$10^{18}$~eV. The obtained results are shown on Fig.~\ref{fig:4} with black
arrows. On the same figure, together with the data from
AGASA~\cite{Shinozaki:apj(2002)}, PAO~\cite{Abraham:app(2008), Aab:jcap(2017)},
and TA~\cite{Rubtsov:pos(2018)}, our previous result~\cite{Glushkov:prd(2010)}
is shown. Curves denote the predictions of theoretical models, developed within
the framework of ``top-down'' scenarios of UHECR
generation~\cite{Ellis:prd(2006), Bhattacharjee:pr(2000),
Berezinsky:plb(1969)}. Shaded area corresponds to the
GZK-mechanism~\cite{Kalashev:jetp(2008)}.

If one considers ``hadronic'' events as a background, then it is possible to
claim with a high degree of confidence, that selected showers were initiated by
primary photons. The probability of at least one such event in our data set
equals to $1 / 959 = 0.001$, and all in the case of all 11 showers it equals to
0.011, which is consistent with other experiments and with results of
theoretical calculations.

\section{Theory of particle acceleration at shocks}
\label{sec:theory}

Theoretical studies of CR origin at the Institute has a long history since the
discovery by G.\,F.~Krymsky of the process of regular (diffusive) acceleration
of charged particles at the fronts of shock waves~\cite{Krymsky:sovfiz(1977)}.
A large amount of theoretical and experimental studies carried out since then
convincingly indicate that the process of regular acceleration underlies a big
number of various phenomena in the generation of spectra of superthermal
particles in various astrophysical objects. Today, almost all studies aimed at
explaining the origin of CR are based on considering the possibilities of
regular acceleration under various conditions. This mechanism turned out to be
so universal and effective that it made it possible to explain many of the
properties of CRs~\cite{Berezhko:ufn(1988)}. Later, a numerical model for study
the diffusive shock acceleration of CRs in supernova remnants has been
developed. It can also describes the evolution of remnants and the properties
of their nonthermal radiation~\cite{Berezhko:app(1994),Berezhko:jetp(1996)}. It
was shown that the acceleration of CR by supernova remnants is characterized by
high efficiency enough to compensate for the losses associated with CRs escape
from the Galaxy. The model also describes well most of the observed properties
of the nonthermal radiation of supernova remnants SN 1006, SN Tycho, SN 1987A
and others.

By applying the established features of CR acceleration in supernova remnants
(magnetic field amplification, CR escape, etc.) to large-scale shock waves
generated in the intergalactic medium by the energy release from active galactic
nuclei, it was shown that nonrelativistic shock waves generated by jets from
active galactic nuclei, can effectively accelerate charged particles up to
energies $E\sim 10^{20}$~eV and able to form the observed CR spectrum of
extremely high energies~\cite{Berezhko:apj(2008)}.

There are two scenarios of the formation of the total CR spectrum are usually
considered. Within the first of them (``dip'' scenario), the galactic CR
component prevails up to energy $10^{17}$~eV, in the energy range
$10^{17}-10^{18}$~eV a transition from the galactic to the extragalactic
component of CRs occurs. In the second scenario (``ankle'' scenario), the
extragalactic component prevails at energies above $10^{19}$ eV, and the
galactic component extended up to $3\times 10^{18}$ eV. The galactic components
in both scenarios are produced in type Ia supernova remnants, where protons are
accelerated up to $3\times 10^{15}$ eV and iron nuclei up to $10^{17}$ eV. In
``ankle'' scenario there is a second component of galactic CRs from more
powerful type IIb supernovae, which explode in the dense wind of a presupernova
star, where maximum energy of accelerated particles can reach $3\times10^{18}$
eV. On Figure \ref{fig:1}d the modeled CR spectra of ``dip'' and ``ankle''
scenarios are shown by solid and dashed lines, respectively. One can see, that
both models describe the experimental data more or less well.

Since the maximum energy of accelerated in supernova remnants CRs is
proportional to the charge of particles, towards the end of the galactic
component heavier elements, up to iron, begin to dominate and the composition
of CRs becomes heavier. On Figures \ref{fig:2}d and \ref{fig:7}b the modeled
value of $\lnA$ as function of energy compared with the modern experimental
data.  Here also ``dip'' and ``ankle'' scenarios are shown by solid and dashed
lines, respectively. It is clearly seen, that the ``ankle'' scenario with the
second galactic component accelerated in type IIb supernova remnants rejected
by existing measurements. Thus, comparison of detailed modeling with
experimental data indicates that galactic CRs are accelerating in supernova
remnants up to $10^{17}$~eV (see~\cite{Berezhko:app(2012)} for more details and
references).

\section{Muon Component of Extensive Air Showers}
\label{sec:muons}

High-energy muons\footnote{with energies $\ge 1$~GeV} are important component
of EASs. They are almost not absorbed in the air and travel through soil and
rock and can be detected deep below the ground level. Since they are direct
descendants of nuclear interactions occurring during air shower development,
the muon density ($\rhom$) is sensitive to the characteristics of said
interactions and also to the mass composition of primary CR initiating EAS in
the Earth's atmosphere.

At the same time, for more than 10 years both experimenters and theorists have
been noting the problem of muon excess in EAS with energies above $10^{17}$~eV
relative to the values obtained in numerical calculations performed with the
use of UHE hadron interaction models. This problem is widely known within the
CR research community as the ``Muon Puzzle''. Recently an international
workgroup has published a meta-analysis that covers a large volume of
experimental data on muon content in EAS~\cite{Dembinski:epj(2019)}. The
analysis evaluates both recent results of currently running experiments
(IceCube~\cite{Gonzalez:epj(2019)}, NEVOD-DECOR~\cite{Bogdanov:yadf(2010),
Bogdanov:app(2018)}, PAO~\cite{Aab:prd(2015), Aab:prl(2016), Muller:epj(2019)}
and {\yeas} among others) and archive data. In the end a preliminary conclusion
was made that none of the hadron interaction models that are widely used in the
UHECR research can adequately describe the behaviour of muon component of EAS
with energies above $10^{18}$~eV. Furthermore, the discrepancy rises with
increasing energy.

Despite the recently obtained hints of a possible solution to the Muon Puzzle
in future experiments at Large Hadron Collider (LHC)~\cite{Albrecht(2021)},
this problems stays unresolved to this day. And besides the discrepancy between
model predictions and experimental data, there is a mismatch between different
experiments.

\subsection{\it The lack of muon excess in Yakutsk data?}

Since registration of EAS is indirect method of CR observation, each array
uses its own experimental technique and utilizes unique methods for
reconstruction of parameters of primary CRs. And each experiment uses its own
classification parameter for estimation of characteristics of the muon
component.

In order to compare muon content in EAS measured on different arrays, authors
of the review~\cite{Dembinski:epj(2019)} have introduced an arbitrary scale
parameter $z$:

\begin{equation}
    z = \frac{\ln{\Nexp} - \ln{\Np}}{\ln{\NFe} - \ln{\Np}}\text{,}
    \label{eq:Z}
\end{equation}
where $\Nexp$ is a classification parameter of muon flux measured in
experiment, $\Np$ and $\NFe$ are, correspondingly, values of this parameters
obtained in simulations for primary protons and iron nuclei.

\begin{figure}[!htbp]
    \centering
    \includegraphics[width=0.95\textwidth]{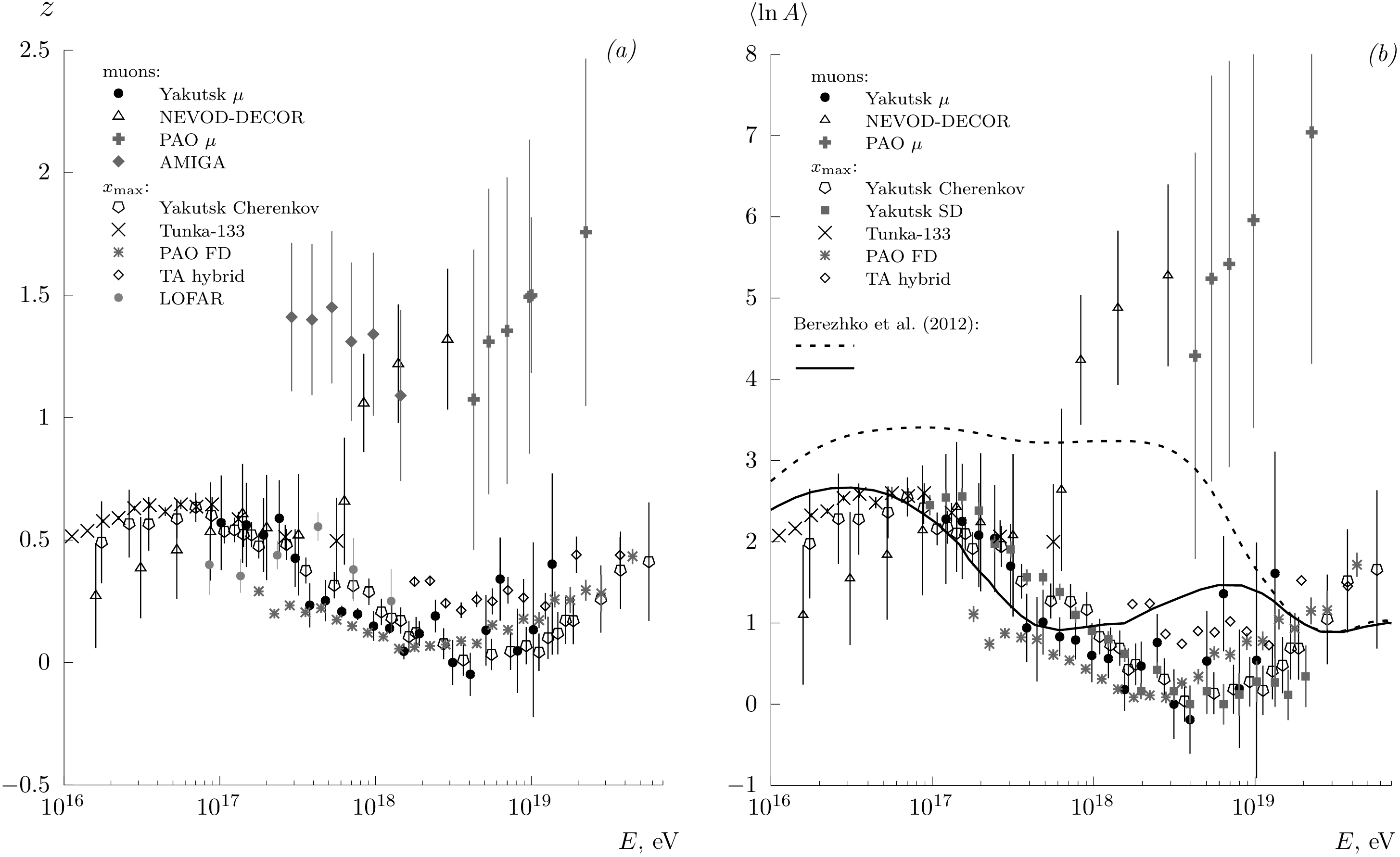}
    \caption{
        (a)\,Values of the scaling parameter $z$ introduced
        in~\cite{Dembinski:epj(2019)}. Only interpretation within the framework
        of the {\qgs}~II-04 model~\cite{Ostapchenko:prd(2011)} are shown since
        all other models demonstrate similar trend. Points denoted as
        ``$\xmax$'' in legend were obtained from the interpretation of the
        longitudinal EAS development results.  Show are data of the
        {\yeas}~\cite{Knurenko:asr(2019)} (empty pentagons),
        Tunka-133~\cite{Prosin:epj(2016)} (crosses), TA~\cite{Abbasi:apj(2018)}
        (empty diamonds), PAO~\cite{Bellido:pos(2018)} (grey asterisks) and
        LOFAR~\cite{Corstanje:2021} (grey circles). Points denoted as ``muons''
        are results of direct measurements of muon fluxes in EAS:
        {\yeas}~\cite{Glushkov:jetpl(2019)} (black circles),
        NEVOD-DECOR~\cite{Bogdanov:yadf(2010), Bogdanov:app(2018)} (empty
        triangles), PAO~\cite{Aab:prd(2015), Aab:prl(2016)} (grey crosses) and
        AMIGA~\cite{Muller:epj(2019)} (grey diamonds).
        (b)\,Estimations of the CR mean atomic number obtained in different
        experiments derived from the data shown in panel (a). Grey squares~---
        estimation derived from the SD LDF~\cite{Pravdin:pazh(2018)}, lines
        represent predictions of the CR diffusive shock acceleration model for
        the case of the ``dip'' (solid line) and ``ankle'' (dashed line)
        scenarios~\cite{Berezhko:app(2012)}.
    }
    \label{fig:5}
\end{figure}

Regular measurements of muon component in air showers at the {\yeas} have
started in 1976~\cite{Glushkov:uhecr(yafsoan1979)}. Current configuration of
the muon setup consists of three detector points located at 0.5 and 1.0~km from
the array center. Each point is an underground housing covered with a
$\sim 2.5$-m layer of soil that forms a shield with $\simeq1$~GeV
threshold and absorbs the e-m component.

At the {\yeas} the main muon estimator is particle density at core distance
300~m~--- $\rhos(300)$~\cite{Glushkov:jetpl(2019)}. The obtained preliminary
values were included in the meta-analysis. Though after applying of energy
rescaling the Yakutsk data became vaguely consistent with the majority of the
presented data, the energy dependence of $z$-parameter differs significantly
from other experiments.

On Fig.~\ref{fig:5}(a) these values are shown without energy rescaling (Yakutsk
$\mu$). It is seen that they are suspiciously consistent with the values
derived from interpretation of the $\xmax$ data of other experiments,
obtained mainly with the use of optical methods and by registering the EAS
radio-emission~\cite{Prosin:epj(2016), Bellido:pos(2018), Abbasi:apj(2018),
Corstanje:2021}. A good agreement is also observed between muon and Cherenkov
data of the {\yeas}~\cite{Knurenko:asr(2019), Glushkov:jetpl(2019)}.

On Fig.~\ref{fig:5}(b) are shown estimations of the CR mean atomic number
obtained in different experiments. It follows that the Yakutsk muon data are
also compatible with estimation derived from the parameters of the LDF measured
by SD~\cite{Pravdin:pazh(2018)}. Both Yakutsk values and estimations based on
the interpretation of the $\xmax$ data of other experiments are also roughly
consistent with predictions of the CR diffusive shock acceleration model for
the case of the ``dip''-scenario described in detail in
Section~\ref{sec:theory} (shown with solid line).

Therefore an intriguing picture is emerging in the UHE domain. Muon
measurements performed at the {\yeas} contradict the results of other
experiments~\cite{Bogdanov:yadf(2010), Bogdanov:app(2018), Aab:prd(2015),
Aab:prl(2016), Muller:epj(2019)} where muon fluxes are measured directly, but
agree with measurements of longitudinal shower development. It is obvious that
it is premature to draw any conclusion, we have to test the treatment and
interpretation of our data at first.

\section{Directional Anisotropy Studies}
\label{sec:anisotropy}

\subsection{\it A beam of ultra-relativistic particles in cosmic rays}

Recently an indication of a short-lived beam of UHE particles was reported.
Three particles with energies above $3 \times 10^{19}$~eV were registered from
a compact sky region during one day: from Jan.\,21, 2009~23:40:35 to Jan.\,22,
2009~22:54:22 (UTC). Two particles were detected at the
{\yeas}~\cite{Artamonov:izvran(1994), Ivanov:msc(2010)} and one~--- at
TA~\cite{Abbasi:apj(2014)}. Chance probability of occurrence of such a triplet
was estimated to be $2.6 \times 10^{-6}$~\cite{Krymsky:pazh(2017)}. Estimated
energies of two particles registered in Yakutsk were nearly identical, $3.63
\times 10^{19}$~eV and $3.55 \times 10^{19}$~eV correspondingly. After the
refinement of the Yakutsk energy estimation~\cite{Glushkov:yadf(2018)} one may
confidently claim that energy of the TA event was twice the energy of two
Yakutsk events~\cite{Krymsky:pazh(2019)}.

\begin{figure}[!htbp]
    \centering
    \includegraphics[width=0.75\textwidth]{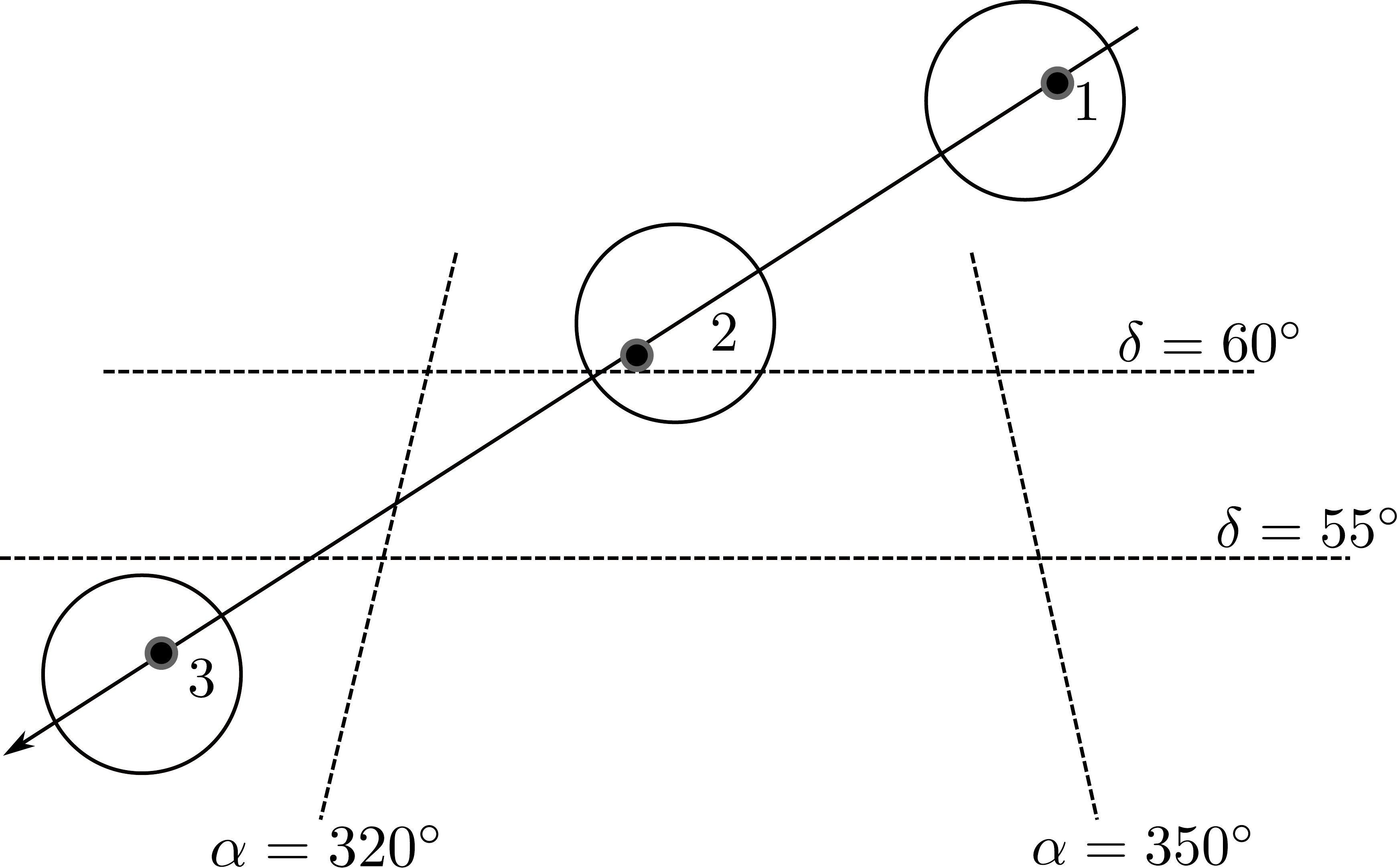}
    \caption{
        A fragment of celestial sphere in equatorial coordinates $(\alpha,
        \delta)$ with three registered events (empty circles). Circle sizes
        roughly correspond to experimental errors. Solid line with an arrowhead
        and black dots represent the best fit to the trajectory of a source
        moving at a constant angular velocity. Particles 1 and 2 were
        registered at the {\yeas}~\cite{Krymsky:pazh(2017)}, particle 3~--- at
        TA~\cite{Abbasi:apj(2014)}.
    }
    \label{fig:6}
\end{figure}

It is only possible to consider all three particles as a beam if they were
emitted by a point source, but their arrival directions lie within $27\degr$
which is inconsistent with this condition. But it is still possible to consider
this hypothesis if the source was moving. Thus, as a possible source of
ultrarelativistic particles a non-stationary relativistic jet was considered.
From the observer's point of view the front of the forming jet may be deemed as
a subluminal point source.

Empty circles in Fig.~\ref{fig:6} indicate the celestial coordinates of all
three events in the equatorial coordinate system. The circle sizes roughly
correspond to the experimental errors. Black line with an arrowhead and black
dots represent the best fit to the parameters under the assumption that the
apparent source of particles was moving with a constant angular velocity. It
is seen that the moving point source hypothesis is actually consistent with
experimental data. If all three events coincided by chance, then such an
ordered arrangement of coordinates would be unlikely. This confirms our
hypothesis that two EAS arrays have registered the arrival of the actual beam of
UHECR particles.

However, since the entire phenomenon lasted for a whole day and this in turn
means that the source was moving much faster then light. This paradox can be
explained by the kinematics of a relativistic source moving toward the
observer. In our case, this effect is amplified by a mechanism that we call the
Cherenkov resonance. The essence of the mechanism is as follows. The source is
a front of a non-stationary relativistic jet that propagates at subluminal
velocity towards the Solar system. Trajectories of UHE particles emitted by the
source are curved in a large-scale magnetic field. The mean particle speed
along the line of sight can become noticeably lower than the speed of light and
lower than the source's speed. Some of the particles emitted by the source on a
certain path segment will then come to the observer simultaneously, in the form
of a beam.

The difference in particle magnetic rigidities should not be great and hence it
follows that one of them could be a helium nucleus and two other particles
could be protons. The phenomenon of the Cherenkov resonance should be general
in nature. In particular, it can be suggested that short gamma-ray bursts also
act as manifestations of the Cherenkov resonance~\cite{Krymsky:pazh(2019)}.

\subsection{\it Zenith angular distribution of cosmic rays}

The zenith angle distribution of EAS from UHECR has been a subject of numerous
studies since the very beginning of EAS measurements. Modern arrays measure
both the sizes and incident directions of showers. Previous studies of zenith
angle distribution $f(\theta, E)$ at the {\yeas} were performed mainly at the
highest energies where the full trigger efficiency is reached. In order to
maximize the statistics and extend the dataset towards the lower energies we
have considered the distribution in the wide energy range, with
energy-dependent array exposure due to absorption of showers in the atmosphere.

The method consists in introducing a unified particle density threshold in the
observed showers and applying of a log-normal distribution for approximation of
fluctuations of the main classification parameter used at the {yeas}~---
$\rhos(600)$. It allows one to describe the expected zenith angular
distribution with a simple analytical formula:

\begin{equation}
    f(\theta) = C \sin(2\theta){\rm ERFC} \left(
        \frac{y_{\text{thr}} - \overline{y(\theta)}}%
        {\sqrt{2} \sigma}
    \right)\text{,}
    \label{eq:ZenDist}
\end{equation}
where $y = \ln\rhos(600)$, $C$ is a normalization constant, $y_{\text{thr}} =
-2.303$, the mean density $\overline{y(\theta)}$ is defined by the attenuation
curve and $\sigma$ is the RMS deviation which is a free parameter during a fit
to experimental data. This allowed us to estimate the particle attenuation
length and the frequency rate of EAS events, both of which are crucial for
obtaining the arrival direction distribution. To obtain the EAS frequency rate
distribution over the zenith angle, within the framework of this approach we
have re-processed our data in the energy range $(10^{17} - 10^{19})$~eV and
studied its tight connection with the $\rhos(600)$ absorption in the atmosphere.

With the use of the resulting distributions we performed a search for the CR
flux anisotropy in equatorial system $(\alpha, \delta)$ and established an
upper limit for a fraction of the CR flux from a point source. The resulting
distribution over the right ascension ($\alpha$) can be described with a
uniform distribution arising from the diurnal rotation of the celestial sphere.

Such a uniform distribution dictates the application of the harmonic analysis.
As a null hypothesis we took an isotropic distribution when expected amplitudes
of all harmonics within the limit $N \rightarrow \infty$ are equal to zero and
a non-zero amplitude of the first harmonic appears if there is a point source
of CR. The phase of the harmonic points the direction towards the source along
the right ascension.

\begin{figure}[!htbp]
    \centering
    \includegraphics[width=0.65\textwidth]{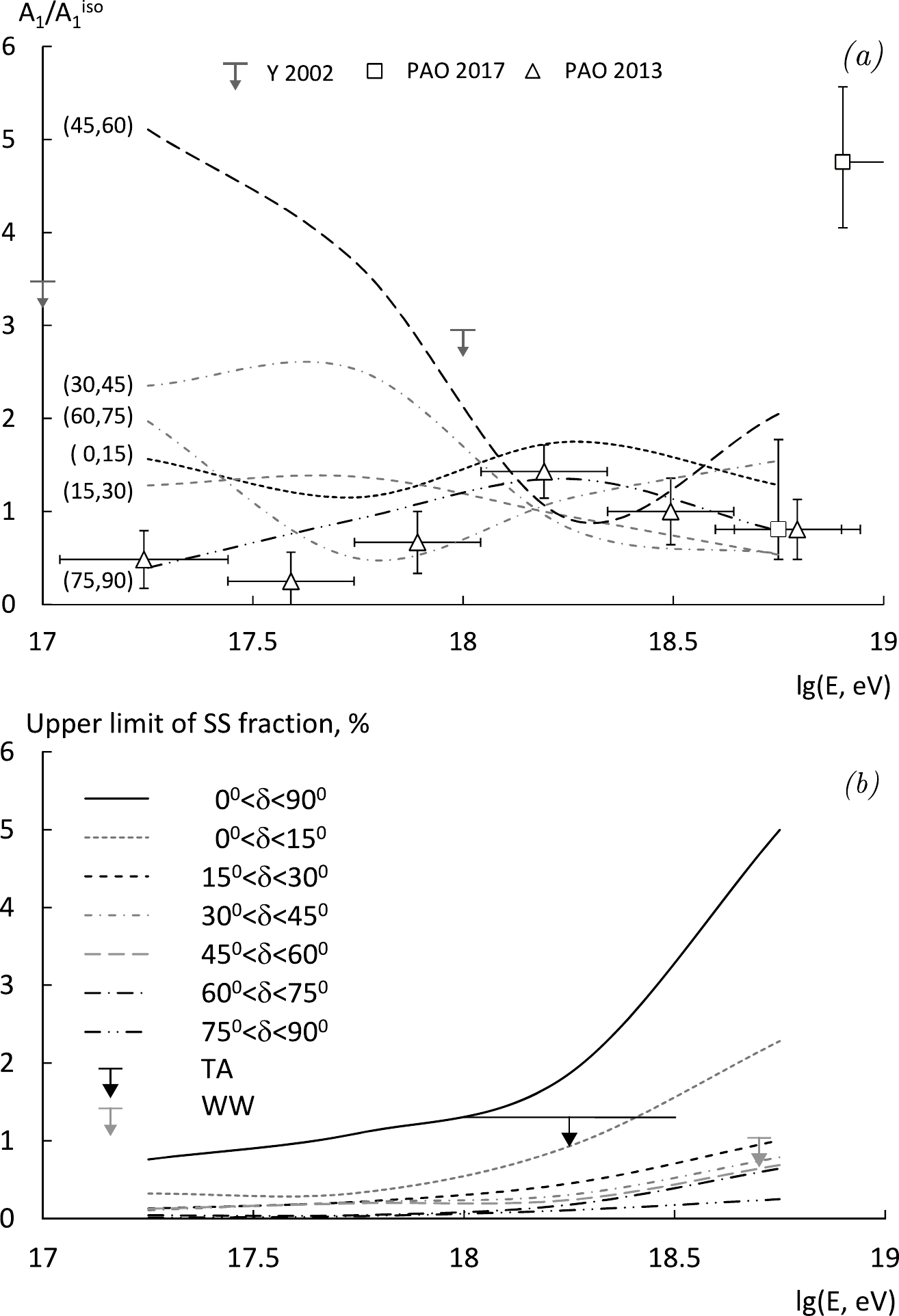}
    \caption{
        (a)\,A comparison of the first harmonic amplitudes of the right
        ascension distribution of CRs measured in the PAO and the {\yeas}
        experiments.  Ratio of observed to expected-for-isotropy amplitudes are
        derived in declination intervals: $-90\degr < \delta < 25\degr$ (PAO
        2013~\cite{Sidelnik:icrc2013}); $-90\degr < \delta < 45\degr$ (PAO
        2017~\cite{Aab:sci(2017)}); $0\degr < \delta < 90\degr$
        (Y~2002~\cite{Pravdin:izvran(2002)}); our latest
        results~\cite{Ivanov:prd(2018)} are denoted by the declination
        intervals $\Delta\delta = 15\degr$.  (b)\,Upper limit of SS fraction in
        declination bins.  TA upper limit on the fraction of EeV protons of
        galactic origin (TA~\cite{Abbasi:app(2017)}) and Wibig \& Wolfendale's
        upper limit on the fraction of galactic light nuclei
        (WW~\cite{Wibig:jpg(1999)}) in the CR beam are given for comparison.
    }
    \label{fig:7}
\end{figure}

To involve the declination distribution in the analysis, one may apply harmonic
analysis withing the rings of $\alpha$ limited in intervals of the declination.
It is convenient to express the first harmonic amplitude in units of isotropic
amplitude which is governed by the sample size $A_1^{\rm iso} =
\sqrt{\pi/N}$. If there is a significant deviation from the expected
value then our null hypothesis should be rejected. The results of this analysis
together with the previous data are shown on Fig.~\ref{fig:7}a.

There is a significant deviation from from isotropy at energies below
$10^{18}$~eV within the declination interval $(45\degr,60\degr)$, but this
effect actually arises from seasonal variations of atmospheric
conditions~\cite{Pravdin:izvran(2002)}. At the same time the data of
PAO~\cite{Aab:sci(2017)} reveal a large-scale anisotropy $A_1 / A_1^{\rm
iso} = 4.8$ at a $5.2\sigma$ confidence level, but the anisotropy dipole points
to the declination $\delta = -24^{+24}_{-13}$ which if outside of the {\yeas}'s
field of view. On the whole, the Yakutsk data do not show a statistically
significant deviation~\cite{Sidelnik:icrc2013} from the null
hypothesis~\cite{Ivanov:prd(2018)}.

There is also a possibility to estimate the statistical power of the method in
a situation when there is a separable source (SS) of CR and we need to
calculate the upper limit for the fraction of the UHECR flux from it in the
total background. Let this SS is presented within the declination interval, at
right ascension $\alpha_{\rm SS}$ and introduces the $f_{\rm SS}$ fraction to
the isotropic flux of CR. Then this alternative hypothesis could be rejected at
a high confidence level by comparing the expected first harmonic amplitude with
the observed one ($A_1^{\rm exp}$).

To calculate the first harmonic amplitude one has to sum $f_{\rm SS}N$ vectors
of the length $2/N$ pointing to a source direction $\alpha_{\rm SS}$, and $(1 -
f_{\rm SS})N$ isotropic vectors of the same length. By applying the central
limit theorem we get the amplitude squared~\cite{Ivanov:prd(2018)}:

\begin{equation}
    \overline{A^2_1} = 4f^2_{\rm SS} + \frac{4(1 - f_{\rm SS})}{N}\text{.}
    \label{eq:SS}
\end{equation}

As a result of comparison with the observed amplitude of the first harmonic of
the right ascension distributions in the declination intervals, the upper
limits of the CR flux fraction from the hypothetical SS, shown on
Fig.~\ref{fig:7}, were obtained in comparison with the previous
results~\cite{Abbasi:app(2017), Wibig:jpg(1999)} in the energy range
$10^{17}-10^{19}$~eV. This result currently represents the strongest constraints
on the fraction of CR flux from a separable CR source in this energy range in
the Northern Hemisphere.

\section{Prospect of the future upgrade}
\label{sec:upgrade}

The {\yeas} is currently being upgraded with the main goals to improve the
accuracy of angular measurements and primary energy estimation. The uncertainty
in energy estimation is expected to be reduced by increasing the accuracy of
particle density measurement in individual events and with introduction of a
new set of Cherenkov light detectors. An upgraded DAQ will provide temporal
measurements with improved accuracy.

For improvement of amplitude measurements a new type of detector station
electronics unit (SEU) was developed, internally nicknamed ``SEU-v.4''. Unlike
the old SEU (``SEU-v.3'') with timing resolution 100~ns and amplitude
RC-resolution 100{\usec}, the SEU-v.4 was developed as an easily
reconfigurable multipurpose device based on a field-programmable gate array
chip (FPGA). It provides the resolution of timing and amplitude RC-measurements
of 6.25~ns.

Synchronization in the new DAQ (internally nicknamed ``EAS-v.4'') is provided
by reconfigurable ``Synchronization Center'' (SC) with a source based on time
server Meinberg M600. Together with SEU-v.4 units it can provide the accuracy
of temporal measurements not worse than $\sim10$~ns which should result
in five-fold increase of the accuracy for reconstruction of EAS arrival
direction. The new system provides online monitoring and real-time diagnostics
of the detectors condition and, what is more important for studying the EAS and
UHECR physics, is capable of performing on-the-fly re-configuration of the
accounting of delays in communication lines. Currently 19 stations located
around the center of the array are equipped with updated SEUs, more stations
are expected to be upgraded in the nearest future. There is also a plan to
implement measurement of the time-base of the detector's signal in the new DAQ.

The main characteristic of differential Cherenkov detectors is timing
resolution: it defines the accuracy of reconstruction of EAS parameters. This
has lead to the design of a high-resolution detector based on the Frensel
lense and Hamamatsu R1250 PMT~\cite{Matarkin:izvran(2019)}. Currently a
prototype cluster of four such detectors undergoes a field-test which has
demonstrated scientific and economic efficiency of such a design. Recently an
algorithm was developed for digital processing of the detector signal time-base
and for reconstruction of saturated pulses. It was shown that the signal
reflects the features of longitudinal development of EAS and is sensitive to
the CR mass composition~\cite{Ivanov:ijmpd(2020)}.

As discussed in Section~\ref{sec:muons}, measurements of muon component are
vital for the EAS studies~--- from both astrophysical and nuclear points of
view. In order to improve the accuracy and statistics of muon measurements at
the {\yeas}, plans are to extend the existing grid of muon detectors.
Preliminary design is to deploy additional ten observational points of 4{\sqrm}
area each and with $\sim 0.7$~Gev threshold.

\section{Conclusion}

The Yakutsk Extensive Air Shower Array has been continuously operating
for more than 50 years and has accumulated a unique and extensive set of data on
various EAS components.

The long noted difference between cosmic-ray fluxes obtained from the array's SD
and its Cherenkov detectors can be mitigated by adjusting both energy estimates
by only 5\,\%, which is well within the energy reconstruction uncertainties of
both methods.

Different methods of measuring the UHECR mass composition give consistent
results. In particular, the mass composition derived from the muon measurements
agrees well with the results of $\xmax$-based methods, both from SD and
Cherenkov detectors.

The possible lack of muon excess according to the Yakutsk data, reviewed in
Section~\ref{sec:muons}, represents an intriguing separate element of the Muon
Puzzle. While presented results contradict to the direct muon measurements
performed by world experiments, they are consistent with both $\xmax$-based
results of other arrays and predictions of theoretical models of UHECR
generation. It is premature to make a final conclusion, first of all it is
necessary to test and re-evaluate the treatment and interpretation of our data
and also to refine the measurements technique. Recently initiated extensive
modernization of the Yakutsk instrument should contribute to solving this
problem.

\begin{acknowledgments}
    This this report was made within the framework of the research project
    No.\,AAAA-A21-121011990011-8 curated by the Ministry of Science and Higher
    Education of the Russian Federation.
\end{acknowledgments}

\bibliography{refs}

\end{document}